\documentclass[pre,twocolumn,superscriptaddress,showkeys,showpacs,amsmath,amssymb,byrevtex,floatfix]{revtex4}

\usepackage[dvips]{graphicx}

\newcommand*{\p}{\mathbf{p}}
\newcommand*{\q}{\mathbf{q}}
\newcommand*{\R}{\mathbf{r}}
\newcommand*{\D}{\mathrm{d}}
\newcommand*{\be}{\begin{equation}}
\newcommand*{\ee}{\end{equation}}

\begin{document}

\title{Hamiltonian dynamics of homopolymer chain models}

\author{Alessandro Mossa}
\affiliation{Department of Chemistry, Rice University, 6100 Main Street,
Houston, Texas 77005, USA}
 
\author{Marco Pettini}
\affiliation{Istituto Nazionale di Astrofisica (INAF), Osservatorio Astrofisico
di Arcetri, Largo E.~Fermi 5, I-50125 Firenze, Italy}

\author{Cecilia Clementi\footnote{Corresponding Author.
Electronic address: cecilia@rice.edu}}
\affiliation{Department of Chemistry, Rice University, 6100 Main Street,
Houston, Texas 77005, USA}

\begin{abstract}
The Hamiltonian dynamics of chains of nonlinearly coupled particles
is numerically investigated in two and three dimensions. 
Simple, off-lattice homopolymer models are used to represent
the interparticle potentials. Time 
averages of observables numerically computed along dynamical trajectories 
are found to reproduce results given by the statistical mechanics of 
homopolymer models.  The dynamical treatment, however, indicates a nontrivial 
transition between regimes of slow and fast phase space mixing.  Such a 
transition is inaccessible to a statistical mechanical treatment and reflects 
a bimodality in the relaxation of time averages to 
corresponding ensemble averages.  It is also found that a change in the 
energy dependence of the largest Lyapunov exponent indicates the 
$\Theta$-transition between filamentary and globular polymer configurations,
clearly detecting the transition even for a finite number of
particles.
\end{abstract}



\date{\today } 

\maketitle

\section{Introduction}
Many important facets of polymer physics can be approached by means of
equilibrium statistical mechanics \cite{DeGennes, VanderzandeBook,RubColby}.  Dynamical 
aspects are typically treated by means of numerical Monte Carlo simulations 
or Langevin dynamics \cite{poly1,poly2,Binder}.
Recent developments (see Ref.~\cite{dyn_rev} for a review and an extensive bibliography) 
in the theory of nonlinear Hamiltonian systems with 
many degrees of freedom provide an alternative viewpoint from which to 
approach polymer physics, and are expected to convey information complementary
to that obtainable by traditional methods.
It is the aim of the present work to use these recent developments to 
investigate the Hamiltonian dynamics of simplified homopolymer models.

The geometrical interpretation of nonlinear Hamiltonian systems with highly 
dimensional phase space has seen considerable progress in the past decade, 
both in terms of theory and in application to physical models 
\cite{MarcoeMonica95,XY2000,Mazzoni}.
Concepts from Riemannian geometry have been used to map 
the natural evolution of a system at constant energy onto geodesic 
trajectories of a manifold in phase space.  This leads to a natural 
definition of Hamiltonian chaos, which is quantified by the largest 
Lyapunov exponent, a measure of how quickly trajectories diverge when 
initially separated by an arbitrarily small distance in phase space.

The study of chaoticity as a function of energy density (energy per degree of 
freedom) is a rich topic with physical applications.  It has been found that in 
many systems there exists a value of the energy density at which a sudden change in 
the energy dependence of the Lyapunov exponent occurs \cite{SST.1,SST.2}. 
For energy densities above 
this value, there is a region of strong chaos, allowing a system to efficiently 
sample phase space, while below this value there is a region of weak chaos, in
which the system cannot readily sample all available phase space.  
Such a transition signifies a strong stochasticity threshold (SST), and is 
associated with a change in the topology of accessible phase space 
\cite{rivNC}. This threshold has also been associated with conventional 
phase transitions \cite{cccp}, and thus provides a novel tool by which to characterize and 
study them. We also note that the efficiency of phase space sampling is 
intimately related to a physical relaxation time, thus providing information
about the time necessary for a system to reach equilibrium.

The ability of Hamiltonian dynamics to quantify the efficiency of phase space
sampling and to detect conventional phase transitions makes it a natural candidate
for the study of polymer dynamics, a field replete with studies using equilibrium 
statistical mechanics and computer simulation methods to characterize 
a polymer phase diagram. 
The Hamiltonian dynamics approach may be particularly useful to provide 
complementary information on the phase space of protein-like polymers,
that present a cooperative transition from a disordered unfolded ensemble to 
a compact folded state.
To complete the folding transition, a protein-like polymer needs to navigate a 
complicated, high-dimensional phase space, that can be populated with
metastable intermediate states and can exhibit a certain degree of local roughness
\cite{Onuchic97,Shea2001,ClementiPlotkin2004,Chavezjacs2004,DasWilsonPNAS2005,dasmatysiakclementi2005,Mazzoni}. 
The purpose of this paper is to show the feasibility and utility of applying
Hamiltonian dynamics to the study of simple homopolymer models as motivation for 
the future study of protein dynamics.

The Hamiltonian dynamics of off-lattice homopolymer models are studied 
at three levels of approximation.  After discussing the general model
in Sec.~\ref{Gen_model}, we consider the free chain approximation in Sec.~\ref{RW_section}, 
the self-avoiding approximation in Sec.~\ref{SAW_model}, and 
the Lennard-Jones model in Sec.~\ref{LJ_model}.
We compare the scaling of the mean squared end-to-end distance computed
along dynamical trajectories with theoretical values computed using
equilibrium statistical mechanics
\cite{DeGennes,DeGennesJPL75,Dupl87,Seno90,ReviewTP,Baum}.  
We also focus on the relaxation time and Lyapunov exponent as a function of energy 
density.  For the Lennard-Jones model, we 
find that the Lyapunov exponent gives a signature of the nonconventional 
$\Theta$-transition between filamentary and globular configurations of the 
homopolymer. This is a strong indication that Hamiltonian dynamics may provide a 
useful tool in characterizing the topology of the accessible phase-space 
of protein-like polymers \cite{protein}.

\medskip
\section {The homopolymer chain model}
\label{Gen_model}

We represent, as it is common in simple models, a polymer as a chain of $N+1$ 
beads connected
by springs, where a configuration of the chain is defined by the positions 
$ \{\q_0,\ldots, \q_N\} $ of the beads in $d$-dimensional continuous space.
The Hamiltonian is defined as
\be
   H=\sum_{i=0}^{N} \frac{\p_{i}^2}{2} + U \,, 
   \label{hamilt}
\ee
where the $\{\p_i \}$ are the canonically conjugate momenta.

A simple choice for the interaction potential~\cite{Baum,ParisieIori1,ParisieIori2} is
\be
\label{potential}
U= \sum_{i=1}^{N} \sum_{j<i} \left[\delta_{i,j+1}  f_\mathrm{spr}(r_{ij})+ \left(\frac{\sigma}{r_{ij}}\right)^{12} -
 \eta_{ij}\left(\frac{\sigma}{r_{ij}}\right)^6 \right],
\ee
where $r_{ij}=|\R_{ij}|=|\q_i - \q_j |$ gives the interparticle distances.
The expression $ f_\mathrm{spr}(r_{ij}) $, representing spring-like anharmonic interactions between 
neighboring  beads in the homopolymer chain, is given by
\be
f_\mathrm{spr}(r_{ij})= \frac{a}{2} (r_{ij}-r_0)^2+ \frac{b}{4} (r_{ij}-r_0)^4 \,,
\ee
where $ r_0 $ is the equilibrium distance between nearest neighbors along the chain.

We can simplify $d$ translational and $d(d-1)/2$ rotational degrees of freedom by 
enforcing the constraints of vanishing total momentum and angular momentum
\be \label{constr}
   \sum_{i=0}^N \p_i=0 \qquad \qquad \sum_{i=0}^N \q_i\times\p_i=0 \,,
\ee
on the initial condition. This way the total energy $E$ is all internal energy and
we can interpret the dynamics generated by the 
Hamiltonian (\ref{hamilt}) as a constant temperature evolution.
 
The Hamiltonian equations of motion,
\be
\dot{\q}_{i}=\frac{\partial H}{\partial \p_i}, \qquad 
\dot{\p}_{i}=-\frac{\partial H}{\partial \q _i}, \qquad
i=0,\ldots,N \: 
\ee
are numerically integrated using an efficient
third order, bilateral, symplectic algorithm \cite{LapoAlgo}. 
The initial conditions are given by random configurations of beads compatible
with the constraints (\ref{constr}). 
 
We expect convergence of relevant observables to ensemble averages after
a transient, nonequilibrated regime. Once equilibrium is attained,
time averages will only fluctuate about ensemble averages.
We are interested in the dependence of the relaxation time $ \tau_\mathrm{R} $ on the
energy per degree of polymerization $\epsilon \equiv E/N$. 

Recall that the ergodic hypothesis underlying
statistical mechanics is not sufficient to ensure the 
convergence of time and ensemble averages in finite time, and that such convergence
requires a stronger condition of a phase space mixing dynamics.
Riemannian geometry provides a useful framework for the study of Hamiltonian 
systems with many degrees of freedom.
This framework has been detailed elsewhere \cite{PhysRep}, and is briefly 
reviewed in App.~\ref{app_geo} of this paper.
A necessary condition for a mixing dynamics is that trajectories in phase space are
unstable with respect to arbitrarily small variations of the initial conditions. 
A quantitative measure of the mean instability of nearby trajectories is given by the 
largest Lyapunov exponent $\lambda_1$ (see App.~\ref{app:lyap} for the definition), 
which also gives a measure of the time scale
necessary for the loss of memory of initial conditions.

It has been shown that regimes of both fast and slow phase space mixing exist 
\cite{SST.1,SST.2,rivNC}.  The crossover between the two is characterized by a sudden
change in relaxation time $\tau_\mathrm{R}$ as a function of energy density $\epsilon$, and 
is intimately connected with a crossover between different scaling laws for 
$\lambda_1(\epsilon)$.
In the language of previous work, this crossover defines a transition from weak to strong
chaos, which governs the efficiency in which phase space is sampled.

\section{The Ideal Polymer}
\label{RW_section}

We initially consider the potential energy $U$ of Eq.~(\ref{potential}) with
the constants $\eta_{ij}$ and $\sigma$ set equal to zero:
\be
   U_\mathrm{RWM}= \sum_{i=1}^{N} \left[ \frac{a}{2}(r_{i,i-1} - r_0 )^2
   + \frac{b}{4}(r_{i,i-1} - r_0 )^4 \right] \,. \label{Hfreechain}
\ee
This defines the simplest idealization of a flexible polymer 
chain, and is the Hamiltonian dynamical counterpart of the random 
walk model (RWM) on a periodic lattice.
Average properties of the simple lattice RWM are easily computed and 
visualized~\cite{DeGennes,VanderzandeBook,RubColby}.

For the ideal homopolymer model in the large $N$ limit, we can analytically compute
the $\epsilon$-dependence of the largest Lyapunov exponent by taking advantage of a
method developed in \cite{modGauss95,modGauss96}. 
The theoretical prediction of $\lambda_1(\epsilon)$, together with other
statistical averages that are analytically obtained for the ideal
model, are then compared with the time averages of the same
quantities numerically computed along the dynamical trajectories. Moreover, in the
dynamical simulations, we also estimate the relaxation time $\tau_\mathrm{R}$
needed for convergence of an observable -- in our case, the mean value of the square
end-to-end distance defined below -- to its expected value.
The dynamical simulations of the system are performed for a finite number of
monomer units, $N$. It is thus also interesting to compare the analytic
results, obtained in the limit $N\rightarrow \infty$, with the numeric results 
obtained at finite values of  $N$.

\subsection{The end-to-end distance}

The size of a polymer chain is usually characterized by its 
\emph{end-to-end distance} $R\equiv|\q_N-\q_0|$. For a RWM, the mean value of $R^2$
can be computed within the canonical ensemble (see App.~\ref{app:avrg} for the details):  
it results to be proportional to $N$
\be
   \langle R^2 \rangle = N\ell^2(\beta) \,.
\ee
The coefficient $\ell$ is a monotonically decreasing function of the inverse 
temperature $\beta=1/(k_\mathrm{B} T)$, whose value tends to $r_0$ as $\beta\to\infty$. 

In the thermodynamic limit, $N\rightarrow \infty$ while $\epsilon$ is constant, we 
can obtain the microcanonical average $\langle A\rangle_{\mu}$ of any observable 
function $A(\{\q_i\})$ in the parametric form \cite{PLV} 
\begin{equation}
\langle A\rangle_{\mu}(\epsilon) \rightarrow \left\{\begin{array}{l}
\langle A\rangle_{\mu}(\beta) = \langle A \rangle^{G}(\beta)\\
\\
\epsilon(\beta) = {\displaystyle
\frac{d}{2\beta} - \frac{1}{N}\frac{\partial}{\partial\beta}
[\log Z(\beta)]} 
\end{array}\right. \,,
\label{mm}
\end{equation}
where $\langle A\rangle^{G}$ is the canonical average of the observable $A$ and 
$Z(\beta)$ is the canonical partition function.
For finite $N$ we have $\langle f\rangle_{\mu}(\beta) = \langle f \rangle^{G}
(\beta )+ {\cal O}(\frac{1}{N})$.

We therefore expect that, for sufficiently large $N$, and after a suitable transition time,
the time averages of the end-to-end distance $ \langle R^2 \rangle $ of the chain, 
numerically 
computed along the dynamical trajectories of the system, will agree with the scaling law
$\langle R^2 \rangle \propto N$ at any fixed value of the energy density $\epsilon$.

%For practical purposes, it is convenient to measure the radius of gyration instead of the 
%end-to-end distance. The radius of gyration of our homopolymer chain is defined as
%\be
%R^2_g = \frac{1}{N+1} \sum_{i=0}^{N}|(\q_i - \q_\mathrm{cm})|^2 \: ,
%\label{eq.radg}
%\ee
%where ${\q}_\mathrm{cm}$ is the center of mass of the system and can be 
%set equal to 0 without
%loss of generality. The mean square value of this quantity is always
%proportional to the mean square value of the end-to-end distance.
%The advantage of using the radius of gyration  rather than the end-to-end distance in 
%numerical calculations is that 
%the former is computed using the coordinates of all particles while the latter uses
%only the coordinates of the endpoints, therefore fluctuations about the mean value 
%are reduced.
 
\begin{figure}
\begin{center}
\includegraphics[width=8.6cm]{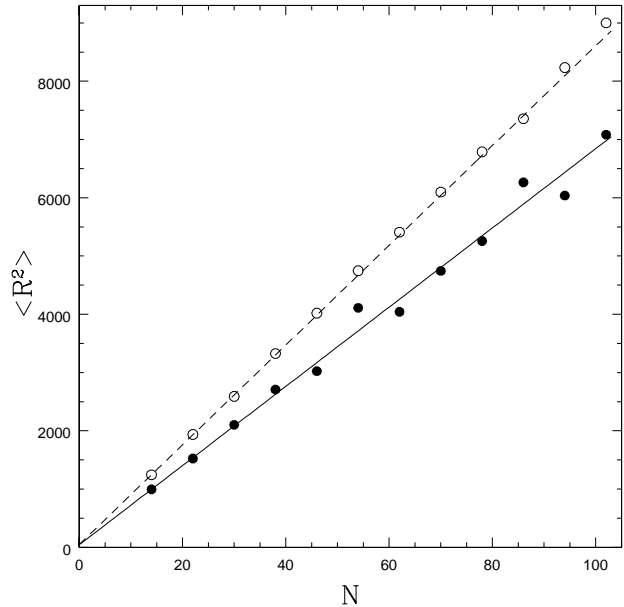}
\caption{\label{fig.rw1} The power law $R^2\propto N$ is correctly recovered
in dynamical averages with a off-lattice random walk model. Results are shown 
for two different values 
of the energy density: $\epsilon=0.1$ (full circles) and $\epsilon=1000$ (open circles).
Solid and dotted line represent the best fit lines, entirely consistent with the 
theoretically predicted behavior.}
\end{center}
\end{figure}

In Fig.~\ref{fig.rw1} we show the comparison between analytical and 
numerical results for two different values \footnote{Here we use 
dimensionless quantities. Physical units can be 
easily introduced when we want to describe a given real system for which 
masses, lengths and parameters of the potentials are assigned.} 
of the energy density $\epsilon$. The
agreement between ensemble and time averages is very good, 
even though the simulated values of $N$ are not very large.

\subsection{The Lyapunov exponent and the relaxation time}
\label{lyfreesection}

\begin{figure}
\begin{center}
\includegraphics[width=8.6cm]{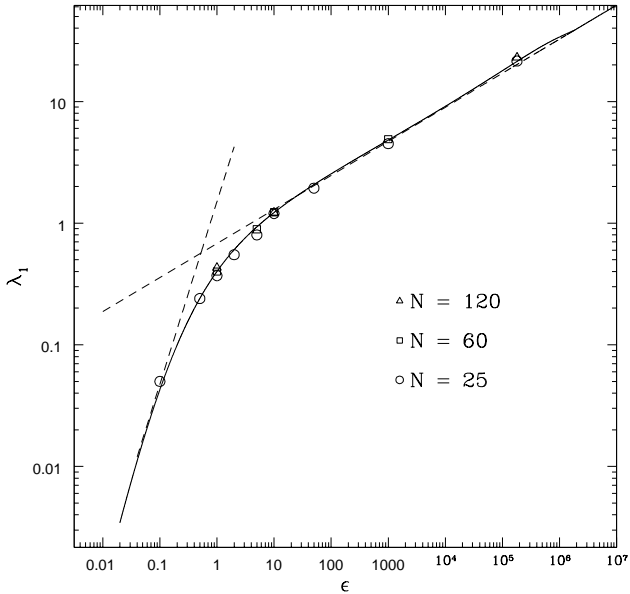}
\caption{Largest Lyapunov exponent $\lambda_{1}$ vs.
energy density $\epsilon$ for the random walk model: comparison between 
theoretical prediction
(solid line) and numerical results (circles); dotted lines are
references to asymptotic power laws $\epsilon^2$ and $\epsilon^{2/3}$, and
their intersection defines the strong stochasticity threshold.}
\label{fig.rwly}
\end{center}
\end{figure}

We now explore the dependence of the largest Lyapunov exponent $\lambda_1$ on the 
energy density $\epsilon$ in order to identify  a crossover in the scaling behavior,
that is the signal of a transition between slow and fast phase space mixing
(i.e., a transition between slow and fast relaxation of time averages to ensemble averages).
A method to analytically compute the largest Lyapunov exponent 
\cite{modGauss95,modGauss96} as a function of $ \epsilon $ is reviewed in 
App.~\ref{app_geo}. 
This method is used to compute $ \lambda_1 $ for the ideal homopolymer chain, 
the results of which are shown in Fig.~\ref{fig.rwly}.

\begin{figure}
\begin{center}
\includegraphics[width=8.6cm]{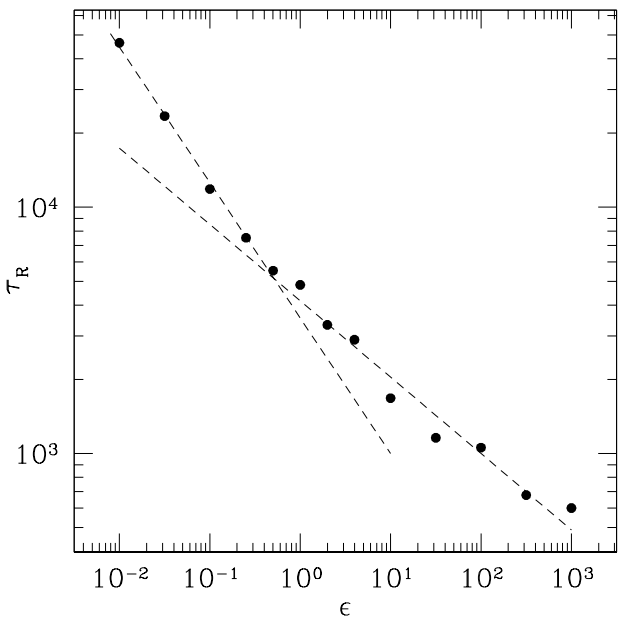}
\caption{Relaxation time $\tau_\mathrm{R}$ (defined as the time of the first occurrence 
of the theoretical value of $\langle R^2\rangle$) vs.~energy density $\epsilon$ for the 
random walk model. The picture shows the results obtained for a chain with $N=22$.}
\label{fig.rwtime}
\end{center}
\end{figure}

Figure \ref{fig.rwly} shows that numerically computed values of $\lambda_{1}$ for 
$N=25$, $N=60$, and $N=120$ are in excellent agreement with analytical results. 
The existence of a crossover energy density $ \epsilon_c $  
between two different scaling laws of the exponent $\lambda_{1}$ is also clear. 
Such a bimodality is characteristic of a SST, 
at which a transition between regimes of a qualitatively different 
chaoticity occurs \cite{SST.1,SST.2}.
The transition from above to below $ \epsilon_{c} $ is coupled with a change in the topology
of phase space trajectories that results in slower phase space mixing \cite{cccp}. 
To further support this,
we have numerically computed $ \tau_\mathrm{R}(\epsilon) $, the relaxation time 
necessary for time averages of the mean square radius $ \langle R^2 \rangle $ to achieve 
theoretically predicted values. 
It is evident from Fig.~\ref{fig.rwtime} that there exists a rather sudden change in the 
behavior of $ \tau_\mathrm{R}(\epsilon) $ at $\epsilon=\epsilon_\mathrm{c}$, 
clear sign of the occurrence of a SST.

\section{The Self-Avoiding Polymer}
\label{SAW_model}

The RWM can be straightforwardly enriched by introducing repulsive interactions 
between monomers. The new potential energy is 
\be \label{HSAW}
   U_\mathrm{SAW}= U_\mathrm{RWM}+ \sum_{i=1}^N \sum_{j<i} 
   \left(  \frac{\sigma}{r_{ij} }  \right) ^{12} \,,
\ee
giving the counterpart to the self-avoiding walk (SAW) on a lattice.  
In lattice models of polymers, short ranged repulsive interactions between
monomers are incorporated through an excluded volume condition, which
requires that any two monomers maintain at least one lattice space separation.
This model, which is more physical than the simple RWM, 
accurately describes the swollen phase of a real polymer chain in a good solvent.
 
The mean squared end-to-end distance for this model scales as $ \langle R^2 
\rangle \propto N^{2 \nu}$.
The existence of a nontrivial exponent ${\nu}$ is known by renormalization group
calculations \cite{Zinn}, but the value, which is expected to depend on dimensionality,
is not analytically known. Flory \cite{Flory} devised a simple approximation scheme 
for computing the exponent ${\nu}$ in any dimension. Basic assumption is that $N$ 
monomers are uniformly distributed within the volume $R^d$ of the swollen polymer: 
all interactions between monomers are disregarded. Given a monomer, the probability 
that another one will fall into its excluded volume $v$ is therefore $Nv/R^d$.
Since there are $N$ monomers, the repulsive energy is 
\be
   \frac{E_\mathrm{rep}}{\kappa T}\sim \frac{N^2v}{R^d}  \,.
\ee
The increase in the size of the polymer comes together with an increase of the entropy,
that for an ideal (i.e. without volume exclusion) chain can be estimated as  
\be
   S\sim \frac{R^2}{N\ell^2} \,.
\ee
The free energy $F=E_\mathrm{rep}-TS$ is minimized by the radius $ R=R_\mathrm{F} $ given by 
\be
   R_\mathrm{F}^{d+2}\sim v\ell^2 N^3 \,,
\ee
which implies that  
\be
   \nu=\frac{3}{2+d} \,.
\ee
In spite of the rather drastic approximations, this is a remarkable assessment, as it gives 
the correct value of $\nu$  for $d=1$ and values within a percent of the most accurate 
numerical results \cite{Zinn} for $d=2$ and $d=3$. However, Flory's theory doesn't work 
when the degree of polymerization $N$ is too small for the approximation of non 
interacting monomers to hold. To be more specific, if the repulsive energy in the 
ideal configuration $R=R_0\sim N^{1/2}\ell$ is 
less than $\kappa T$, the polymer doesn't swell, so that $\nu=1/2$ as in the RWM case. 
The repulsive interactions are relevant only when the so-called chain interaction parameter \cite{RubColby} 
\be \label{zdef}
   z\sim \frac{v}{\ell^d}N^\frac{4-d}{2}  
\ee
is large enough, $z \gg 1$. The above formula implies that for $d\ge 4$ the effect of 
repulsion becomes negligible and the polymer may be considered ideal. 

In the following, we use the framework of Flory's theory to interpret the scaling 
behavior of the off-lattice SAW for different values of the ratio $\sigma/r_0$ in 
a unified way. 

\subsection{Scaling behavior of $\langle R^2 \rangle$}
\label{SAW_section}
The repulsive potential $\left(\frac{\sigma}{r_{ij}}\right)^{12}$ in the 
Hamiltonian (\ref{HSAW}) is sufficiently steep that we can assume two monomers 
$i$ and $j$ cannot get closer than a distance $r_{ij}\approx \sigma$. The 
excluded volume is therefore $v\approx\sigma^d$, while the equilibrium distance 
between two nearest neighbors along the chain is given by $\ell\approx r_0$. 
In view of Flory's theory, we expect to observe a linear dependence on $N$ of
$\langle R^2\rangle$ when 
\be
   N\le N_\mathrm{c}\sim\left(\frac{\sigma}{r_0}\right)^{-\frac{2d}{4-d}} \,,
\ee 
while for $N\ge N_\mathrm{c}$ we should observe $\langle R^2\rangle\propto N^{2\nu}$.

\begin{figure}
\begin{center}
\includegraphics[width=8.6cm]{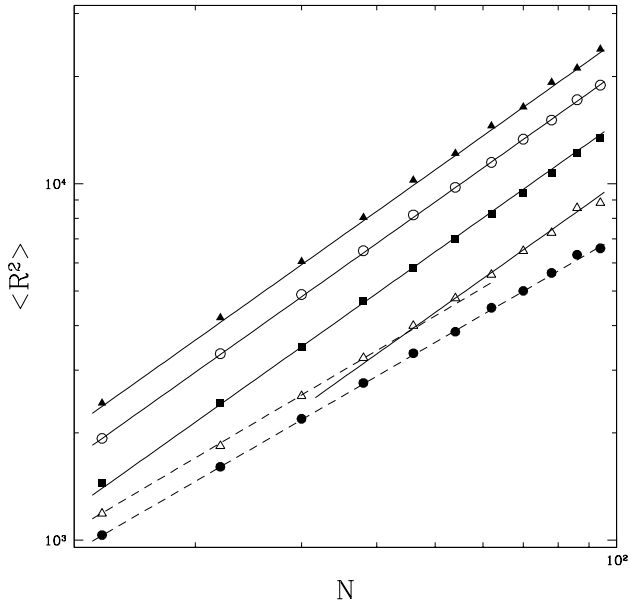}
\caption{Mean squared end-to-end distance for different values of the ratio $\sigma/r_0$ (from bottom to top, 0.1, 0.3, 0.5, 0.7, 0.9) for the model defined by Eq.~(\ref{HSAW}). The solid lines follow the power law of the 3-dimensional self-avoiding walk $N^{2\nu}$, the dashed lines represents the behavior linear in $N$ of the random walk model.}
\label{fig.saw1}
\end{center}
\end{figure}

\begin{figure}
\begin{center}
\includegraphics[width=8.6cm]{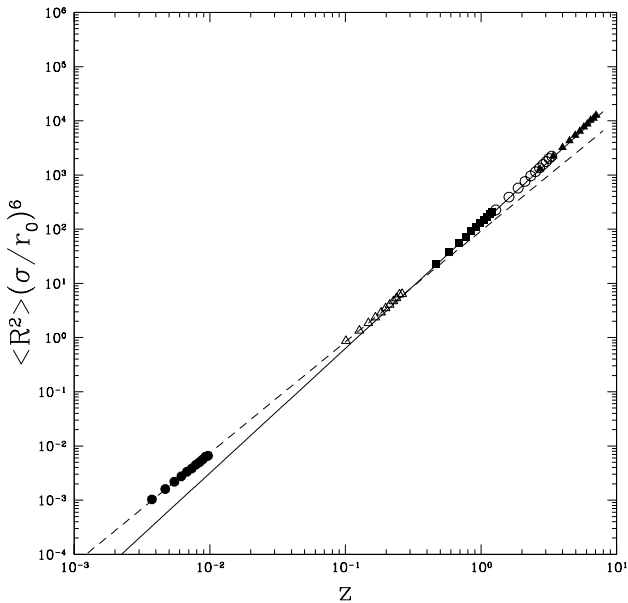}
\caption{Collapse plot for the data of Fig.~(\ref{fig.saw1}). The dotted and solid line 
are the two asymptotic behaviors in Eq.~(\ref{funzia}).}
\label{fig.sawscal}
\end{center}
\end{figure}

Figure \ref{fig.saw1} presents the results of numerical investigation of the 
dynamical behavior of self-avoiding homopolymers containing
between $10$ and $100$ monomers for several values of $ \frac{\sigma}{r_0} $.
The analysis of data shows that:
\begin{itemize}
  \item $\frac{\sigma}{r_0} \le 0.1$ gives $ \langle R^2 \rangle \propto N$ (the value extracted from the best fit is $0.99\pm0.02$),
        as expected from the RWM;
  \item $\frac{\sigma}{r_0} \ge 0.5$ gives $ \langle R^2 \rangle \propto 
        N^{\frac{6}{5}} $ (numerical values from the three linear fits are between 1.19 and 1.21 with errors of order 0.02), which is the expected SAW scaling;
  \item $\frac{\sigma}{r_0} = 0.3$ can't be satisfactorily fitted with just one line: a change in the scaling of $\langle R^2 \rangle$ occurs at an intermediate value of $N$.
\end{itemize}
We assume that the information in Fig.~(\ref{fig.saw1}) can be synthesized as follows:
\be
   \left\{\begin{array}{rcl} 
      \langle R^2 \rangle & = & k_1 N \qquad \ \ \, N<N_c \\
      \langle R^2 \rangle & = & k_2 N^{2\nu} \qquad N>N_c
   \end{array} \right.   
\ee
where 
\be
   N_c=\alpha\left(\frac{\sigma}{r_0}\right)^{-\frac{2d}{4-d}} \,,
\ee
with the constant $\alpha$ independent on $\sigma$ or $r_0$. Continuity for $N=N_c$ dictates the following relation between $k_1$ and $k_2$:
\be
   \frac{k_1}{k_2}=\alpha^\frac{4-d}{2+d}\left(\frac{\sigma}{r_0}\right)^{-\frac{2d}{2+d}} \,.
\ee
We can now express $N$ in terms of the parameter $z$ defined by Eq.~(\ref{zdef}), thus obtaining the scaling relation
\be
   \frac{\langle R^2 \rangle}{k_1}\left(\frac{\sigma}{r_0}\right)^\frac{2d}{4-d}=f(z) \,,
\ee
where
\be \label{funzia}
   f(z)=\left\{\begin{array}{lcl} 
      z^\frac{2}{4-d} &\mathrm{for}& z<\alpha^\frac{4-d}{2} \\
      \alpha^\frac{d-4}{d+2}z^\frac{12}{(d+2)(4-d)} &\mathrm{for}& z>\alpha^\frac{4-d}{2}
   \end{array} \right. \,.
\ee
Within the approximations of Flory's theory, we can therefore treat in a unified way 
the transition between the random and the self-avoiding walk behavior. As a consistency
check, Fig.~\ref{fig.sawscal} shows how the curves of Fig.~\ref{fig.saw1} collapse onto a
unique curve $f(z)$ once expressed in terms of $z$. 

\subsection{Dynamical behavior}
\begin{figure}[t]
\begin{center}
\includegraphics[width=8.6cm]{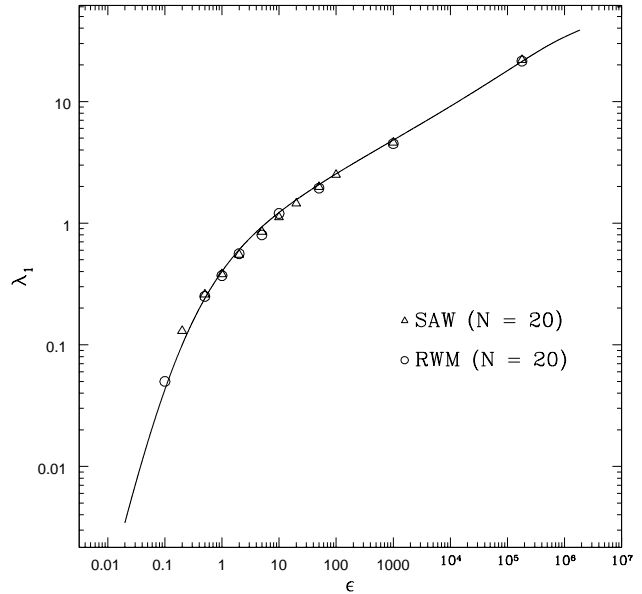}
\caption{Comparison between the largest Lyapunov exponent for the random walk
case (circles) and the self-avoiding walk case (triangles). The solid line is the
analytic prediction for the random walk case.}
\label{fig.sawly}
\end{center}
\end{figure}

The behavior of the Lyapunov exponent of the self-avoiding chain is  investigated 
in this section. Figure~\ref{fig.sawly} compares 
the numerical~\cite{algBenettin} results for the SAW ($ \sigma/r_0=\sqrt[12]{10^{11}} $) 
and the analytical results for the RWM ($ \sigma=0 $).  It is clear that 
the analytic curve computed with $ \sigma=0 $ provides a good approximation for 
the numerically computed Lyapunov exponent as a function of 
energy density, even when $ \sigma/r_0=\sqrt[12]{10^{11}} $.  Even though the scaling law
changes from $ \langle R^2 \rangle \sim N$ for the RWM to 
$ \langle R^2 \rangle \sim N^{\frac{6}{5}} $ for the SAW, the dynamics, as characterized 
by $ \lambda_1 (\epsilon) $, maintains features of the random walk.  That is, the 
introduction of the repulsive energy in the Hamiltonian has no apparent effect on
the instability of dynamical trajectories in phase space.  

This fact can be explained as follows:
Because the repulsive potential grows very steeply for $ r_{ij} < 1 $, the phase space for 
the SAW can be thought of as that for the RWM with inaccessible regions.
If these forbidden regions of phase space are not too large ($\sigma/r_0<1$), we expect 
that sufficiently uniform observables in phase space will 
not be significantly affected.  In the strongly
chaotic regime, the average Ricci curvature and its fluctuations, 
which determine the value of Lyapunov exponent (see App.~\ref{app_geo}), do not change 
significantly when measured
in different regions of the ambient manifold.  As such, they are not sensitive to the 
exclusion of some regions of phase space, and thus we expect the Lyapunov exponent
to be negligibly affected.  We expect that this reasoning may cease to be valid when 
the range of the repulsive potential approaches the interatomic distance ($\sigma/r_0\ge 1$).

\section{The Lennard-Jones polymer}
\label{LJ_model}

The results from previous sections have demonstrated that the application of 
Hamiltonian dynamics to the study of homopolymers is both feasible and provides
dynamical insight otherwise inaccessible.  We now consider a homopolymer chain
in which attractive forces are present in addition to the repulsive forces 
considered in the previous section.  Recalling our potential (\ref{potential}),
we set $\eta_{ij} = \eta$, where $\eta$ is a constant, for all $i$ and $j$.  
For more complicated heteropolymers such as minimalist protein models, the values of $\eta_{ij}$
can be selected to effectively represent the interactions between different 
amino-acid types (see, \textit{e.g.}, \cite{dasmatysiakclementi2005}).
This, however, is beyond the scope of our current work and is left for future study.  

The intermonomer potential (\ref{potential}) can be rewritten in the form
\be \label{ULJ}
   U_\mathrm{LJ}=U_\mathrm{RWM}+\sum_{i=1}^N\sum_{j<i} 4\gamma \left[\left(
   \frac{\lambda}{r}\right)^{12}-\left(\frac{\lambda}{r}
   \right)^{6}\right] \,,
\ee
where $\gamma= \eta^2/4$ and $\lambda=\sigma/\sqrt[6]{\eta}$.

It is known that a homopolymer chain with this potential can adopt two 
distinct phases associated with the dominance of either the attractive or repulsive 
interaction energy \cite{Flory}.  With the Lennard-Jones potential above, 
the relative importance of the attractive and repulsive energies can be varied by 
varying the temperature.  At high temperature, the attractive part of the potential 
is negligible and the chain exists in a swollen phase, closely mimicking the 
SAW model in which only repulsive forces are relevant. 
The power law of the SAW, $\langle R^{2} \rangle \propto N^{2 \nu}$, is recovered. 
At low temperatures, the attractive term becomes relevant and the chain collapses
into a compact configuration, resulting in the radius of the system scaling as 
$\langle R^2 \rangle \sim N^{2/d}$. The separation between the self-avoiding and 
collapsed chain 
regimes is marked by a $\Theta$-temperature which indicates a phase transition.
We study the properties of the Lennard-Jones homopolymer model in dimension $d=2$, 
and thus expect Flory's estimate of $\nu=3/4$ to hold for the SAW, and $\nu=1/2$ to hold
for the collapsed chain.

\subsection{Scaling laws}

\begin{figure}
\begin{center}
\includegraphics[width=8.6cm]{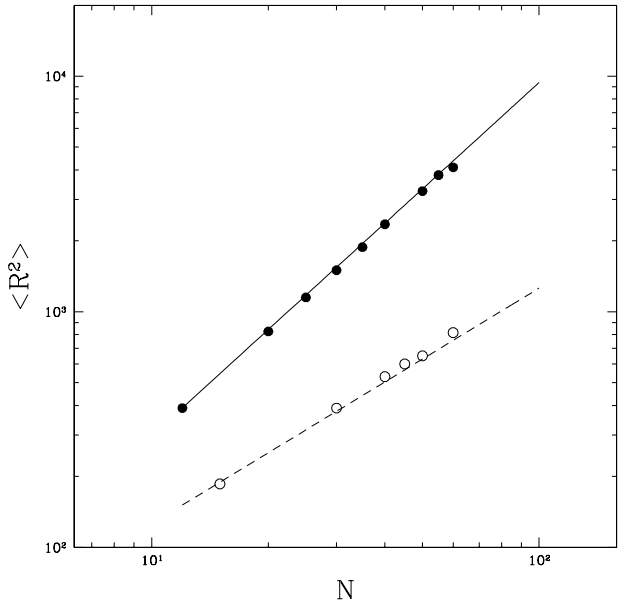}
\caption{The Lennard-Jones polymer (\ref{ULJ}) in dimension 2 exhibits different scaling 
laws above (full circles, $\epsilon=10$) and below (open circles, $\epsilon=-2$) 
the $\Theta$-point. Solid line is reference to $\langle R^2\rangle\propto N^{3/2}$, dotted 
line to $\langle R^2\rangle\propto N$. }
\label{fig.lj1}
\end{center}
\end{figure}

Dynamical simulations are performed in the range $N\in [10,100]$. We fix the parameters 
$\gamma$ and $\lambda$ so that at high temperature the system gives the scaling
of Section \ref{SAW_section} for this range of $N$.
In the dynamical simulations we vary the energy density $\epsilon$,
which is a function of the temperature according to Eq.~(\ref{mm}), and in our
microcanonical simulations acquires the meaning of mean energy per
degree of freedom.  Figure~\ref{fig.lj1} shows that two different scaling laws 
are found for values of the energy density $\epsilon=10$ and $\epsilon=-2$,
implying that at these values of $\epsilon$ the system is in different phases.
In fact, the value of the exponent jumps between $\nu=3/4$ and $\nu=1/2$
when the energy density reaches a value $\epsilon_\theta$. This transition has 
been proven to be a tricritical point \cite{DeGennes}.

The proper way to estimate the value $\epsilon_\theta$ at which the 
transition occurs, as well as the tricritical exponent $\nu_\theta$, would be 
to perform a finite size scaling analysis along the lines illustrated in 
Ref.~\cite{Flavio88}. Unfortunately, we cannot follow this route because the 
fluctuations of the end-to-end distance
in our off-lattice model are much more pronounced compared to the lattice SAW 
models usually employed for this measure. A rough estimate of $\epsilon_\theta$
can be obtained nevertheless by studying the variation of the exponent $\nu$ with 
respect to the energy density. With the choice of parameters $a=0.1, b=10, \lambda=9,
\gamma=2$, the tricritical exponent $\nu_\theta=4/7$ 
\cite{Dupl87} corresponds to an energy density between $-0.8$ and $-0.7$ (the yellow 
line in Fig.~\ref{fig.ljly.2}).   

%In order to estimate the critical value $\epsilon_{\theta}$
%at which the transition occurs, we have performed a finite size scaling 
%analysis. An immediate way of analyzing approximate results for $\langle R^2\rangle_{N}$, at
%finite $N$ and near a $\Theta$-transition, consists of computing effective
%exponents \cite{fss} in the form
%\be
%2 \nu_N = \frac{\ln( \frac{R^2_N}{R^2_{N-1}} ) } {\ln\left(\frac{N}{N-1}\right)} \,.
%\ee
%The curves at different $N$ have a clear tendency to intersect each
%other in a restricted region, as showed in Fig.~\ref{fig.ljscal}.
%Abscissas and ordinates of these intersection points yield approximate values 
%of $\epsilon_{\theta}$ and $\nu_{\theta}$, respectively~\cite{Flavio88}. 
%
%\begin{figure}
%\begin{center}
%\includegraphics[width=8cm]{fig8.ps}
%\caption{{Plot of the behavior of $\nu(\epsilon)$ for different 
%values of $N$.}}
%\label{fig.ljscal}
%\end{center}
%\end{figure}
%

\subsection{The Lyapunov exponent and dynamical characterization 
of the $\Theta$-transition}

It has been recently conjectured that, in the presence of a second order phase 
transition, there exists a close relationship between dynamical properties 
described by Lyapunov exponents and statistical ensemble averages 
~\cite{delpo1,delpo2,IO,Lando,cccp}.  In particular, the behavior of the Lyapunov
exponent is expected to change abruptly as a function of energy density $\epsilon$ 
near the transition.  This provides motivation to seek whether there is a dynamical
signature marking the nonconventional $\Theta$-transition expected in our 
homopolymer system.  

\begin{figure}
\begin{center}
\includegraphics[width=8.6cm]{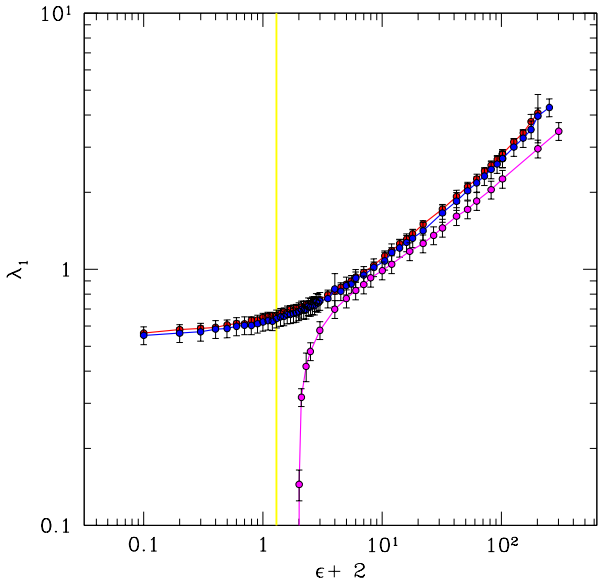}
\caption{Lyapunov exponent as a function of the energy density for $N=50$ (blue) and 
$N=100$ (red) Lennard-Jones chains. The purple points show the behavior of a $N=50$
SAW, while the purple line is the theoretically predicted value for an infinite length
SAW. The yellow line marks the estimated value of $\epsilon_\theta$.}
\label{fig.ljly.2}
\end{center}
\end{figure}

To test for a dynamical indication of the $\Theta$-transition, we have numerically
computed the largest Lyapunov exponent $\lambda_1(\epsilon)$ of the system 
as a function of energy density $\epsilon$.  A sharp change in the slope of
$\lambda_1(\epsilon)$ takes place for a value of the energy density close to the 
transition value $\epsilon_{\theta}$ estimated above, as can be seen
in Fig.~\ref{fig.ljly.2}.  Even though the $\Theta$-transition is only properly defined
in the thermodynamic limit $N \rightarrow \infty$, numerical calculations of the 
Lyapunov exponent for $N=50$ (blue points) and $N=100$ (red points) already appear to 
indicate the existence of a phase transition.
  
%\begin{figure}
%\begin{center}
%\includegraphics[width=8cm]{fig08.ps}
%\caption{Cross-over between two different behaviors of the Lyapunov
%exponent at the $\Theta$-point for the 2-dimensional Lennard-Jones 
%homopolymer.}
%\label{fig.ljly}
%\end{center}
%\end{figure}

We have analytically computed the high temperature behavior of $\lambda_1(\epsilon)$
in order to validate the numerical calculations by comparing 
to the behavior of observables in the limit $N\rightarrow \infty$. 
Even though the statistical averages are not exactly computable, we 
use the method described in App.~\ref{app_geo}, in which
the attractive part of the potential is negligible and the chain is,
in practice, a self-avoiding one. 
We note that the behavior of the Lyapunov exponent for the self-avoiding polymer in 
three dimensions is equivalent to the random walk.

A similar result holds for the self-avoiding chain in two dimensions, but a more 
careful estimate is necessary to compute the average quantities described 
in App.~\ref{app_geo}.
This is because the integrals diverge when integrating over all possible 
configurations using the canonical measure $\exp(-\beta H)$ with the 
free chain potential $U_\mathrm{RWM}$ (\ref{Hfreechain}).
Physically, however, it is very unlikely that two particles will get closer 
than the length scale $\gamma$ of the Lennard-Jones interaction.
Thus, we can approximate the difference between the canonical measure for the 
free chain and the canonical measure for the self-avoiding chain by 
neglecting a spherical volume of radius $\gamma$ around each
particle. Such a procedure does not change the result in three
dimensional space because the integrand functions vanish in
these regions. 
In Fig.~\ref{fig.ljly.2} we show the analytical result for the Lyapunov 
exponent obtained by applying this method. The agreement with the corresponding
numerical values is very good  at energy density greater than the critical value 
$\epsilon_{\theta}$, which corresponds to the high temperature phase. 
At lower energy density, the attractive part of the Lennard-Jones potential ceases 
to be negligible, and agreement with the values for the SAW case is lost.

\section{Conclusions}

We have shown that the Hamiltonian dynamics description of continuous space 
homopolymer chain models reproduces the statistically predicted scaling laws, 
even for homopolymer chains with a small number of monomer units.
This result is nontrivial since the dynamical model allows monomer units
to move freely in continuous space, with changes in conformation being 
driven by nontrivial differentiable and conservative dynamics.  
The dynamical description of homopolymer systems also conveys qualitatively 
new information, giving a physical time scale, $\tau_\mathrm{R}$, that measures the amount
of time necessary for time averages to converge to their statistical 
counterparts for a system with arbitrary initial conditions.
  Thus, if $\tau_\mathrm{R}$ were to exceed the observational time lapse, 
experimental measurements 
would be inadequate to predict statistical mechanical averages. 
Moreover, $\tau_\mathrm{R}$ has nontrivial energy density dependence, and 
a sudden change of $\tau_\mathrm{R}(\epsilon )$ 
is observed to correspond with the strong stochasticity threshold that 
exists for non-integrable Hamiltonian systems with many degrees of freedom.
This transition, signaled by a crossover in the 
$\epsilon$-dependence of the largest Lyapunov exponent, occurs between
dynamical regimes which exhibit qualitatively different 
efficiency of phase space mixing.

The dynamical approach to homopolymer dynamics also demonstrates a clear 
change in the scaling of the radius of gyration with $N$ as the 
self-avoidness parameter in the potential was considered in addition
to the simple nearest neighbor potential that maintains connectivity.  It 
was seen that the scaling law depends on the relative magnitude
of the effective range of the repulsive potential and of the equilibrium monomer
spacing, which is physically expected.  Previous methods to determine scaling
relied on computing the radius of gyration for various particle numbers, and thus the 
dynamical approach provides and alternative method by which to study scaling relations.

The dynamical method applied to homopolymers also provides a clear signature 
of the $\Theta$-transition between filamentary and globular configurations.  
These results are obtained for relatively small $N$, while 
the statistical mechanical approach relies on the thermodynamic limit, $N 
\rightarrow \infty$.  The results are physically relevant, especially for 
polymers containing small numbers of monomer units.  

We anticipate that the extension of this paper's techniques of Hamiltonian dynamics 
to more complex heteropolymer systems such as minimalist protein models (see, \textit{e.g.}, \cite{dasmatysiakclementi2005,Mazzoni})
will provide useful information to complement current statistical mechanical
and simulation studies.
The dynamical signature of a transition from fast to slow phase space mixing in
a homopolymer model suggests that a similar dynamical signature indicating the
transition from strong to weak chaos may exist in protein-like systems. 

\begin{acknowledgments}
This work was supported by grants from NSF (Career CHE-0349303, CCF-0523908 and 
CNS-0454333), ATP (003604-0010-2003), the Robert A. Welch Foundation (Norman
Hackermann Young Investigator award, and C-1570).

We acknowledge Steve Abel for his contribution to this project. We thank 
members of Clementi's groups for stimulating discussions.
\end{acknowledgments}

\appendix

\section{Statistical average of $\langle R^2\rangle$ for a RWM} \label{app:avrg}

Statistical averages are computed in the canonical ensemble, giving 
the partition function for the system in $ d $ dimensions as 
\be \label{partfunc}
Z=\int \prod_{i=0}^N \D \p_{i}\int \prod_{i=0}^N \D \q_i \, \exp[-\beta H(\q,\p)]
\ee
where the Hamiltonian is given by Eq.~(\ref{hamilt}) 
with potential energy~(\ref{Hfreechain}).
The integration over the momenta is straightforward and gives a factor
\be
  Z_p=\left(\frac{2\pi}{\beta}\right)^{\frac{N+1}{2}d} \,.
\ee
The remaining integration is simplified upon a change in the variables from the 
$\{\q_i\}$ to the $\{\R_i\}\equiv\{\R_{i,i-1}\}$ (the Jacobian of this transformation 
is $(N+1)^{-d}$), so that 
\be
Z_q=(N+1)^{-d}\left(\frac{2\pi^{\frac{d}{2}}}{\Gamma(d/2)}\right)^{N} \mathcal{I}_d^{N} \,,
\ee
where $\Gamma$ is the Euler function and
\be \label{integralI}
\mathcal{I}_d\equiv \int_{0}^{+\infty}\D r\, r^{d-1}
 \exp \left[ -\frac{a \beta}{2} (r-r_0)^2 -\frac{b \beta}{4} (r-r_0)^4 \right] \,.
\ee
The square of the distance from one end of the polymer chain to the other can be written 
\begin{eqnarray}
R^2 &=& | {\q}_N - {\q}_0 | ^2 = \sum_{i,j=1}^{N}\R_i \cdot \R_j
\end{eqnarray}
Hence for the mean value of the end-to-end distance we find 
\begin{eqnarray}
 \langle R^2 \rangle &  = &  \frac{\int \prod_{i=1}^{N} \D\R_i \sum_{i,j=1}^{N}\R_i\cdot\R_j \,
  \exp(-\beta H)}{\int \prod_{i=1}^{N} \D\R_i \exp(-\beta H)}  \nonumber \\
  & = & \frac{\int \prod_{i=1}^{N} \D\R_i \sum_{i=1}^{N}|\R_i|^2 \,
  \exp(-\beta H)}{\int \prod_{i=1}^{N} \D\R_i \exp(-\beta H)}  \nonumber \\
  & = & N \frac{\mathcal{I}_{d+2}}{\mathcal{I}_d} \equiv N\ell^2(\beta) \,.
\end{eqnarray}
Not surprisingly, the power law observed in the RWM on a lattice is recovered. 
The coefficient $\ell(\beta)$ represents the mean distance between 
particles in the chain (counterpart to the lattice constant in the discrete model). 
By evaluating the integral Eq.~(\ref{integralI}), we easily see that $\ell(\beta)$  
is a monotonically decreasing function of $\beta$, going to $r_0$ in the limit 
$\beta\to\infty$.

\section{Definition of the largest Lyapunov exponent} \label{app:lyap}

Given that $\dot x^i = X^i(x^1\dots x^n)$ is a dynamical system,
and if we denote by $\dot\xi^i = {\cal J}^i_k[ x(t)]\, \xi^k$
the usual tangent dynamics equation, where
$[{\cal J}^i_k]$ is the Jacobian matrix of $[X^i]$, then the largest Lyapunov
exponent $\lambda_1$ is defined by
\begin{equation}
\lambda_1 = {\displaystyle\lim_{t\rightarrow\infty}}~\frac{1}{t}\ln
\frac
{\Vert\xi (t)\Vert}{\Vert\xi (0)\Vert} \; .
\label{eq3}
\end{equation}
By setting $\Lambda [x(t),\xi (t)]= \xi^T\,{\cal J}[x(t)]\,
\xi /\,\xi^T\xi\equiv\xi^T {\dot\xi} /\xi^T \xi
=\frac{1}{2}\frac{d}{dt}\ln (\xi^T\xi )$,
this can be formally expressed as a time average
\begin{equation}
\lambda_1 ={\displaystyle\lim_{t\rightarrow\infty}}~\frac{1}{2t}\int_0^t
\,d\tau \,\Lambda [x(\tau ), \xi (\tau )] \;.
\end{equation}

\section{Riemannian theory of Hamiltonian chaos}
\label{app_geo}

\subsection{General framework}

We provide a brief summary of the concepts involved in a
differential geometric method to describe Hamiltonian chaos and to
theoretically predict the value of the largest Lyapunov exponent.
The geometrical formulation of the dynamics of conservative 
systems was first used by Krylov in his studies of the dynamical
foundations of statistical mechanics \cite{Krylov} and subsequently
became a standard tool in the study of abstract systems like Anosov flows
in the framework of ergodic theory.
In recent papers \cite{modGauss95,modGauss96},
the geometric approach has been successfully extended and applied to explain 
the origin of chaos in models of physical relevance.

Consider a system with $N$ degrees of freedom defined by the Lagrangian 
${\cal L} = T - V$, in which the kinetic energy is quadratic in the velocities,
$T=\frac{1}{2}a_{ij}\dot{x}^i\dot{x}^j$.  Such dynamics can be recast in 
geometrical terms since the natural motions are the extrema of either
the Hamiltonian action functional, ${\cal S}_\mathrm{H} \equiv \int{\cal L} \mathrm{d}t$,
or the Maupertuis action, ${\cal S}_\mathrm{M} \equiv 2 \int T\, \mathrm{d}t$.
In fact, the geodesics of a Riemannian manifold are themselves the extrema of the
arc-length functional $\ell=\int \sqrt{g_{ij}\mathrm{d}x^i \mathrm{d}x^j}$.
Hence a suitable choice of the metric tensor allows
identification of the arc length with either ${\cal S}_\mathrm{H}$ or
${\cal  S}_\mathrm{M}$, and of the geodesics with the natural motions of the
dynamical system. 

Starting from ${\cal  S}_\mathrm{M}$ we obtain
the Jacobi metric on the accessible configuration space,
$(g_J)_{ij} = [E - V(\{x\})]\,a_{ij}$.  The extrema 
of the Hamiltonian action ${\cal S}_\mathrm{H}$ can be described as geodesics 
of a manifold using Eisenhart's metric on an enlarged configuration space-time,
$\{t\equiv x^0,x^1,\ldots,x^N,x^{N+1}\}$, where real coordinate $x^{N+1}$
is related to the action \cite{Marco93}.  The arc-length is
\begin{equation}
ds^2 = -2V(\textbf{x}) (dx^0)^2 + a_{ij} dx^i dx^j + 2 dx^0
dx^{N+1}~.
\label{ds2E}
\end{equation}
The manifold has a Lorentzian structure and the dynamical
trajectories are geodesics satisfying the condition
$ds^2 = C \mathrm{d}t^2$, where $C$ is a positive constant.
In the geometrical framework, the stability
of trajectories is mapped to the stability
of geodesics, and is thus completely determined by the
curvature properties of the underlying manifold.
The field $J$, commonly known as the Jacobi field, 
obeys Jacobi equation %\cite{Marco93}
\begin{equation}
\nabla_{\dot\gamma}^2 J + R(\dot\gamma,J)\dot\gamma = 0~,
\label{eqJ}
\end{equation}
where $\nabla_{\dot\gamma}$ is the covariant derivative
along the geodesic $\gamma (s)$, $R$ is the Riemann curvature tensor, and
$\dot\gamma$ is the velocity field along $\gamma$.
Thus $J$ is seen to measure the deviation between nearby geodesics. 

In the case of isotropic, hyperbolic manifolds, the curvature
term in Eq. (\ref{eqJ}) can be rewritten as $R(\dot{\gamma},J)\dot{\gamma} = KJ$, 
where the sectional curvature $K$ is a negative constant, 
thus giving that equation (\ref{eqJ}) has exponentially
growing solutions and that the system is dynamically unstable. This is the origin 
of chaotic dynamics in Anosov flows.
The geometric picture for coupled nonlinear oscillators, however, is much 
different since all curvatures (Ricci, scalar, sectional) are mainly
(and in some cases strictly) positive. 
Even so, exponentially growing solutions of the stability equation (\ref{eqJ}) can
be obtained through parametric resonance even if no
negative curvature is experienced by the geodesics
\cite{Marco93,MarcoeLapo93,modGauss95,modGauss96}.
In the large $N$ limit and under the assumption that the manifold is
nearly isotropic, this mechanism can be modeled by
replacing Eq. (\ref{eqJ}) with an effective scalar
Jacobi equation \cite{modGauss95,modGauss96} which reads
\begin{equation}
\frac{d^2\psi}{ds^2} + \kappa(s) \, \psi = 0~,
\label{eqpsi}
\end{equation}
where $\psi^2 \propto
|J|^2$, and the ``effective curvature''
$\kappa(s)$ is a stochastic, $\delta$-correlated gaussian process.
The mean $\kappa_0$ and variance $\sigma_\kappa$ of $\kappa(s)$
are identified respectively with the average and the root-mean-square (RMS)
fluctuations of the Ricci curvatures at any given point $k_R = K_R/N$
(which is itself the average of the sectional
curvature over the directions of $J$) along a geodesic,
\begin{equation}
\kappa_0 =   \frac{\langle K_\mathrm{R} \rangle}{N}~,   \qquad
\sigma^2_\kappa  =
\frac{\langle (K_\mathrm{R} - \langle K_\mathrm{R} \rangle)^2 \rangle}{N}~.
\label{kappa}
\end{equation}

Using Eisenhart's metric
$K_\mathrm{R}= \Delta V = \sum_{i=1}^N \partial^2 V/\partial x_i^2$, the exponential
growth rate $\lambda_1$ of the envelope of solutions to Eq. (\ref{eqpsi}),
which provides a natural estimate of the Lyapunov exponent, 
can be computed exactly, with
\begin{equation}
\lambda = \frac{\Lambda}{2} - \frac{2 \kappa_0}{3 \Lambda}\,,\qquad
\Lambda = \left(2\sigma_\kappa^2 \tau +
\sqrt{\frac{64 \kappa_0^3}{27} + 4\sigma_\kappa^4 \tau^2}~\right)^\frac{1}{3} . 
\label{lambda}
\end{equation}
Here
$2\tau = (\pi\sqrt{\kappa_0})/(2\sqrt{\kappa_0 (\kappa_0 +
\sigma_\kappa}) + \pi\sigma_\kappa)$, and in the limit 
$\sigma_\kappa/\kappa_0 \ll 1$, one finds $\lambda \propto \sigma_\kappa^2$. 
The details can be found in Refs.~\cite{modGauss95,modGauss96}.

\subsection{Lyapunov exponent for an ideal homopolymer}

We use Eisenhart's metric to compute the mean and RMS fluctuations of the 
Ricci curvature, 
\be
K_\mathrm{R} = \sum_{i=0}^N \frac{\partial ^2 V}{\partial x_i^2},
\ee
which are necessary to determine the largest Lyapunov exponent, $\lambda_1$.
In $d$ dimensions, the Hamiltonian described by Eq.~(\ref{hamilt}) with potential 
energy~(\ref{Hfreechain}) gives
\begin{widetext}
  \begin{eqnarray*}
    K_\mathrm{R} &=& \sum_{j=0}^N \sum_{\mu=1}^d \frac {\partial ^2 V(q)}{\partial
    q_{i,\mu}^2} \\  
    & = & 
    2\sum_{j=1}^N\left( a+3b(r_{j,j-1}-r_0)^2+(d-1)\frac{a(r_{j,j-1}-r_0)+b(r_{j,j-1}-r_0)^3}{r_{j,j-1}} \right) \,. 
  \end{eqnarray*}
\end{widetext}
In order to compute the Gibbs average, we modify the
canonical partition function Eq.~(\ref{partfunc}) such that
\be \label{ztilde}
\tilde{Z}(\alpha)=\int\prod_{i=0}^N \mathrm{d}{\p}_{i}\int\prod_{j=0}^N\mathrm{d}\textbf{q}_j \  
\exp \left[-\beta H({\p},\textbf{q})+\alpha K_\mathrm{R} \right] \, .
\ee
The problem of computing the mean and RMS values of the curvature 
thus reduces to taking derivatives of the modified partition function (\ref{ztilde}), since
\be
\frac{\langle K_\mathrm{R}\rangle}{N} =\frac{1}{N}\left[\frac{\partial}{\partial \alpha}
\ln \tilde{Z}(\alpha)\right]_{\alpha=0}
\ee
and
\be
\frac{\langle (K_\mathrm{R}-\langle K_\mathrm{R} \rangle)^2\rangle}{N}=\frac{1}{N}
\left[\frac{\partial^2}{\partial \alpha^2}
\ln \tilde{Z}(\alpha)\right]_{\alpha=0} \,.
\ee
The factorization of the partition function along the lines of App.~\ref{app:avrg} yields
\be
\tilde{Z}(\alpha)=\left(\frac{2\pi}{\beta}\right)^{\frac{N+1}{2}d}(N+1)^{-d}\left(
   \frac{2\pi^{\frac{d}{2}}}{\Gamma(d/2)}\right)^N\tilde{\mathcal{I}}^N_d(\alpha)
\ee
where
\begin{widetext}
\be
\tilde{\mathcal{I}}_d(\alpha)= 
\int_{0}^{+\infty} \mathrm{d}r\, r^{d-1}
 \exp \left[ -\frac{a \beta}{2} (r-r_0)^2 -\frac{b \beta}{4} (r-r_0)^4+\alpha k(r) \right] \,,
\label{intfreek}
\ee
\end{widetext}
and
\be
k(r)=2 a d - 2 a (d-1) \frac{r_0}{r} + 6 b (r-r_0)^2 + 2 b (d-1)
\frac{(r-r_0)^3}{r}.
\ee
Thus the quantities to be computed are the one dimensional integral in 
Eq. (\ref{intfreek}) and its derivative with respect to $\alpha$, 
each of which is readily computed numerically.


\begin{thebibliography}{99}

\bibitem{DeGennes}
P.-G.~de~Gennes,
\textit{Scaling Concepts in Polymer Physics}
(Cornell University Press, Ithaca, NY, 1979).

\bibitem{VanderzandeBook}
C.~Vanderzande, \textit{Lattice Models of Polymers} (Cambridge University
Press, Cambridge, 1998).

\bibitem{RubColby}
M.~Rubinstein and R.~H.~Colby, 
\textit{Polymer Physics} (Oxford University Press, Oxford, 2003). 

\bibitem{poly1}
\textit{Computer Simulation in Chemical Physics}, edited by M.~P.~Allen and
D.~J.~Tildesley, NATO ASI Series C, Vol. \textbf{397}, (Kluwer, Dordrecht, 1993).

\bibitem{poly2}
K.~Kremer, Macromol. Chem. Phys. \textbf{204}, 257 (2003).

\bibitem{Binder} K.~Binder, J. Phys.: Condens. Matter \textbf{16}, S429 (2004).

\bibitem{dyn_rev} M.~Pettini \textit{et al.}, Chaos \textbf{15}, 015106 (2005).

\bibitem{MarcoeMonica95}
M.~Cerruti-Sola and M.~Pettini,
Phys. Rev. E \textbf{51}, 53 (1995).

\bibitem{XY2000}
M.~Cerruti-Sola, C.~Clementi, and M.~Pettini, Phys. Rev E \textbf{61}, 5171 (2000). 

\bibitem{Mazzoni}
L.~Mazzoni and L.~Casetti, cond-mat/0603409.

\bibitem{Onuchic97}
J.N.~Onuchic, Z.~Luthey-Schulten, P.G.~Wolynes, 
Annu Rev Phys Chem \textbf{48}, 545 (1997).

\bibitem{Shea2001}
J.-E.~Shea, and C.~Brooks~{III}, 
Annu Rev Phys Chem \textbf{52}, 499 (2001).

\bibitem{ClementiPlotkin2004}
C.~Clementi and S.S.~Plotkin, 
Protein Sci. \textbf{13}, 1750 (2004).

\bibitem{Chavezjacs2004}
L.L.~Chavez, J.N.~Onuchic, and C.~Clementi, 
J Am Chem Soc \textbf{126}, 8426 (2004).

\bibitem{DasWilsonPNAS2005}
P.~Das, C.J.~Wilson, G.~Fossati, P.~Wittung-Stafshede, K.S.~Matthews, and C.~Clementi, 
Proc. Natl Acad. Sci. USA \textbf{102}, 14569 (2005).

\bibitem{dasmatysiakclementi2005}
P.~Das, S.~Matysiak, and C.~Clementi,
Proc. Natl Acad. Sci. USA \textbf{102}, 10141 (2005).

\bibitem{SST.1}
M.~Pettini and M.~Landolfi,
Phys. Rev. A \textbf{41}, 768 (1990).

\bibitem{SST.2}
M.~Pettini and M.~Cerruti-Sola,
Phys. Rev. A \textbf{44}, 975 (1991).

\bibitem{rivNC} L.~Casetti \textit{et al.}, Riv. Nuovo Cim. \textbf{22}, 1 (1999).

\bibitem{cccp}
L.~Caiani, L.~Casetti, C.~Clementi, and M.~Pettini,
Phys. Rev. Lett \textbf {79}, 4361 (1997).

\bibitem{DeGennesJPL75}
P.-G.~de~Gennes,
J. Physique Lett. \textbf{36} L55, (1975); \textbf{39} L299, (1978).

\bibitem{Dupl87}
B.~Duplantier and H.~Saleur,
Phys. Rev. Lett. \textbf{59}, 539 (1987).

\bibitem{Seno90}
C.~Vanderzande, F.~Seno, and A.~L.~Stella,
Phys. Rev. Lett. \textbf{61}, 1520 (1988).

\bibitem{ReviewTP}
K.~De~Bell and T.~Lookman,
Rev. Mod. Phys. \textbf{65}, 87 (1993).

\bibitem{Baum}
A.~Baumg\"{a}rtner,
J. Chem. Phys. \textbf{72}, 871 (1980).

\bibitem{protein} A first attempt to analyze protein folding pathways in the
light of chaotic instability of the dynamics has been provided in:
M. Braxenthaler \textit{et al.}, PROTEINS \textbf{29}, 417 (1997).

\bibitem{ParisieIori1}
G.~Iori, E.~Marinari, and G.~Parisi,
J. Phys: Math. Gen. A \textbf{24}, 5349 (1992).

\bibitem{ParisieIori2}
G.~Iori, E.~Marinari, G.~Parisi, and M.~V.~Struglia,
Physica A \textbf{185}, 98 (1992).

\bibitem{LapoAlgo}
L.~Casetti, 
Physica Scripta,  \textbf {51}, 29 (1995).

\bibitem{PhysRep} L.~Casetti, M.~Pettini, and E.~G.~D.~Cohen,
Phys. Rep. \textbf{337}, 237 (2000).

\bibitem{Marco93}
M.~Pettini,
Phys. Rev E \textbf {47}, 828 (1993).

\bibitem{MarcoeLapo93}
L.~Casetti and M.~Pettini,
Phys. Rev E \textbf{48}, 4320 (1993).

\bibitem{modGauss95}
L. Casetti, R. Livi, and M. Pettini,
Phys. Rev. Lett. \textbf{74}, 375 (1995).

\bibitem{modGauss96}
L. Casetti, C. Clementi, and M. Pettini,
Phys. Rev. E \textbf{54}, 5969 (1996).

\bibitem{PLV}
J. L. Lebowitz, J. K. Percus, and L. Verlet,
Phys. Rev. \textbf{153}, 250 (1967).

\bibitem{Zinn} 
J.~C.~Le Guillou and J.~Zinn-Justin, 
Phys. Rev. Lett. \textbf{39}, 95 (1977);
Phys. Rev. B \textbf{21}, 3976 (1980)  

\bibitem{Flory}
P.~Flory,
\textit{Principles of Polymer Chemistry}
(Cornell University Press, Ithaca, NY, 1971).

\bibitem{algBenettin}
G.~Benettin, L.~Galgani, and J.~M.~Strelcyn,
Phys. Rev. A \textbf{14}, 2338 (1976).

\bibitem{fss}
B.~Derrida and H.~Saleur, J. Phys. A \textbf{18}, L1075 (1985); 
V.~Privman, J. Phys. A \textbf{19}, 3287 (1986).

\bibitem{Flavio88}
F.~Seno and A.~L.~Stella,
J. Phys. France \textbf {49}, 739 (1988).

\bibitem{delpo1}
C.~Dellago, H.~A.~Posch, and W.~G.~Hoover,
Phys. Rev E \textbf {53}, 1485 (1996).

\bibitem{delpo2}
C.~Dellago and H.~A.~Posch,
Physica A \textbf {230}, 364 (1996); \textbf {237}, 95 (1997); \textbf {240}, 68 (1997).

\bibitem{IO}
C.~Clementi, 
Laurea Thesis in Physics, Universit\`a di Firenze, Italy (1995).

\bibitem{Lando}
L.~Caiani, Laurea Thesis in Physics, Universit\`a di Firenze, Italy (1995).

\bibitem{Forest}
E. Forest and R. D. Ruth,
Physica D \textbf {43}, 105 (1990).

\bibitem{Krylov}
N. S. Krylov,
\textit {Works on the foundations of statistical mechanics}
(Princeton University Press, Princeton, NJ, 1979).

\end{thebibliography}
\end{document}